\documentclass[11pt,a4paper]{article}
\usepackage{jheppub}
\usepackage{amsmath}
\usepackage{multicol,bbm}
\usepackage{enumerate}
\usepackage{bibentry}
\usepackage[toc,page]{appendix}

\graphicspath{{./Figs/}}

\newcommand{\beqs}{\begin{equation*}}
\def\beq{\begin{equation}}

\newcommand{\eeqs}{\end{equation*}}
\def\eeq{\end{equation}}

\def\beq{\begin{equation}}
\def\eeq{\end{equation}}
\def\be{\begin{equation}}
\def\be{\begin{equation}}
\def\ee{\end{equation}}
\def\ba{\begin{eqnarray}}
\def\ea{\end{eqnarray}}
\def\bea{\begin{eqnarray}}
\def\eea{\end{eqnarray}}
\def\eq{\begin{equation}}
\def\eqe{\end{equation}}
\def\eqa{\begin{eqnarray}}
\def\eqae{\end{eqnarray}}

\def\beqa{\begin{eqnarray}}
\def\eeqa{\end{eqnarray}}

\def\de{\partial}

\title{Exploring soft constraints on effective actions}
\author[a]{ Massimo Bianchi,}
\author[a]{Andrea L. Guerrieri,}
\author[b]{ Yu-tin Huang,}
\author[b]{Chao-Jung Lee,}
\author[a]{Congkao Wen}
\affiliation[a]{Dipartimento di Fisica, Universit\`a di Roma ``Tor Vergata"  I.N.F.N. Sezione di Roma ``Tor Vergata", Via della Ricerca Scientifica, 00133 Roma, Italy}
\affiliation[b]{Department of Physics and Astronomy, National Taiwan University, Taipei 10617, Taiwan, ROC} 
\emailAdd{massimo.bianchi@roma2.infn.it}
\emailAdd{andrea.guerrieri@roma2.infn.it}
\emailAdd{yutinyt@gmail.com}
\emailAdd{r02222074@ntu.edu.tw}

\emailAdd{Congkao.Wen@roma2.infn.it} 

\abstract{We study effective actions for simultaneous breaking of space-time and internal symmetries. Novel features arise due to the mixing of Goldstone modes under the broken symmetries which, in contrast to the usual Adler's zero, leads to non-vanishing soft limits. Such scenarios are common for spontaneously broken SCFT's. We explicitly test these soft theorems for $\mathcal{N}=4$ sYM in the Coulomb branch both perturbatively and non-perturbatively. We explore the soft constraints systematically utilizing recursion relations. In the pure dilaton sector of a general CFT, we show that all amplitudes up to order $s^{n} \sim \partial^{2n}$ are completely determined in terms of the $k$-point amplitudes at order $s^k$ with $k \leq n$. Terms with at most one derivative acting on each dilaton insertion are completely fixed and coincide with those appearing in the conformal DBI, {\it i.e.} DBI in AdS. With maximal supersymmetry, the effective actions are further constrained, leading to new non-renormalization theorems. In particular, the effective action is fixed up to eight derivatives in terms of just one unknown four-point coefficient and one more coefficient for ten-derivative terms. Finally, we also study the interplay between scale and conformal invariance in this context.}
\begin{document}

\maketitle 

\section{Introduction}  

Effective actions in general contain an infinite number of higher dimensional operators whose precise coefficients require detailed understanding of their ultra-violet (UV) completion. In particular, except for low energy global symmetries and some positivity constraints~\cite{Adams:2006sv}, these coefficients are in principle arbitrary. On the other hand for effective theories associated with spontaneous symmetry breaking, it has long been known that soft theorems associated with the broken symmetries can be exploited to constrain the S-matrix, and in turn the effective action. Famous examples include Adler's zero for single U(1) Goldstone boson (GB)~\cite{Adler:1964um}, as well as its non-abelian extension~\cite{Low:2014nga}. Recently it has been shown that a class of effective field theories, including non-linear sigma models, Dirac Born-Infeld (DBI) and a special Galileon, can be completely determined through the use soft theorems~\cite{Cheung:2015ota}.

When spacetime, or both spacetime and internal symmetries are spontaneously broken, the soft-limits of GB's in general will no-longer vanish and are proportional to lower point amplitudes.\footnote{Flat space DBI action has vanishing soft limits due to the vanishing of amplitudes with odd number of external legs.} This is due to the fact that there are multiple GB's that mix under the broken symmetries. That this is true can be understood from the Ward identity of the broken generator:
\eq\label{Master}
\partial_\mu\langle J^\mu(x)\phi(x_1)\cdots\phi(x_{n{-}1})\rangle=-\sum_{i=1}^{n{-}1} \delta(x-x_i)\langle \phi(x_1)\cdots\delta\phi(x_i)\cdots\phi(x_{n{-}1})\rangle\,.
\eqe
If $\delta\phi$ leads to a state in the physical spectrum, then the RHS can lead to a non-vanishing result upon LSZ reduction and thus a non-vanishing soft limit. The conventional vanishing soft-pion limits simply reflect the fact that pions shift under the broken symmetry, and hence $\delta\phi$ does not lead to a physical state under infinitesimal transformations.

For broken conformal symmetry, the Goldstone modes that arise from dilatation and conformal boost are not independent, leading to a single dilaton~\cite{Low:2001bw}. This implies that the soft-dilaton limit can be non-vanishing, as the broken symmetries relate the dilaton to itself. Indeed the plurality of broken generators is reflected in the universality of the single soft dilaton behaviour. In particular expanding the $n$-pt amplitude involving one dilaton in terms of its soft momentum leads to leading and sub-leading terms that are simply proportional to the $(n{-}1)$-point amplitude~\cite{Boels:2015pta, DiVecchia:2015jaq}. In the presence of other global symmetries, the broken generators can rotate the dilaton into the new GB's and {\it vice versa}. This is a common situation for super conformal field theories on the Coulomb or Higgs branch, where both conformal and R-symmetry are broken. Consider for example $D=4$, $\mathcal{N}=4$ super Yang-Mills ({S}YM) in the Coulomb branch, where the massless scalars comprise one dilaton and 5 GB's for R-symmetry breaking SO(6)$\rightarrow$SO(5). As the broken R-symmetry generators mix the GB's and the dilaton, we will find non-vanishing soft limits. In this perspective, the Coulomb branch effective action of maximal {S}CFTs not only enjoys maximal supersymmetry but also exposes ``maximal broken symmetry".

Note that these soft theorems must be respected both in the UV where massive degrees of freedom are present, and in the infrared (IR) where they are integrated out. In this paper we verify this perturbatively by computing the one-loop effective action of $\mathcal{N}=4$ {S}YM up to six fields. This is done by considering the one-loop amplitude of maximal SYM in higher dimensions with the extra component of loop momenta identified as the mass of the massive multiplet. Expanding the integrand around the large mass limit, the integral yields the matrix element of the effective action. For non-perturbative tests, we examine the amplitudes from the instanton effective action obtained in~\cite{Bianchi:2015cta}. We have verified the validity of the new soft theorems to order $s^5$ at six points and $s^{10}$ at five points for one-loop amplitudes, where $s$ generically denotes Mandelstam invariants $s_{ij}=2k_i{{\cdot}}k_j$. While for the amplitudes generated from the one-instanon effective action~\cite{Bianchi:2015cta} are always of order $s^4$ for the scalar sector, we have confirmed the soft theorems for pure-dilaton amplitudes to nine points and for dilaton and pion mixed amplitudes up to seven points. In~\cite{Huang:2015sla,Luo:2015tat} leading and sub-leading soft theorems have also been checked against the amplitudes generated by the dilaton effective action, related to the trace anomaly in the recent study of the $a$-theorem~\cite{Schwimmer:2010za,Komargodski:2011vj,Elvang:2012st,Elvang:2012yc,Dymarsky:2013pqa,Schwimmer:2013jma,Bobev:2013vta}.

Soft theorems provide additional information on the analytic structure of scattering amplitudes, which can be combined with factorization constraints to recursively construct higher multiplicity results.  Armed with the dilaton soft theorems, one can show that the matrix elements of the pure dilaton effective action are fully determined by a subset of operators via on-shell recursion~\cite{Luo:2015tat}. In particular, at $2n$-derivative order, the S-matrix for any multiplicity, {\it i.e.} any number of dilaton insertions, is completely determined in terms of operators of the form ${\partial}^{2k}\varphi^k$ for $k\leq n$.  For maximal susy, the dilaton effective action for arbitrary number of dilatons are fixed up to ten derivatives in terms of three parameters: the coefficients of four-point operators at orders $s^2, s^4$ and $s^5$. For $D=4$, $\mathcal{N}=4$, we find that the dilaton amplitude  at $s^{2}$ and $s^3$ are one and two-loop exact respectively for arbitrary multiplicity. At orders $s^{4}$ and $s^5$, amplitudes with arbitrary multiplicity are completely determined in terms of the four-point coefficient. Beyond $s^5$ higher point coefficients are necessary to determine the $n$-point amplitude. 

Dilaton soft theorem is separated in two pieces, reflecting the fact that there are two kinds of generators being broken, scale and conformal boost. A theory endowed with only scale invariance will satisfy the leading soft theorem but not the sub-leading one. Thus the question of scale vs conformal symmetry becomes to which extent sub-leading soft theorem follows from leading. We study this question beginning with five-point amplitudes to very high order in $s$ (until $s^{11}$), and show that amplitudes satisfying the leading soft theorems automatically satisfy sub-leading soft theorem. Similar statements hold if one considers the amplitudes determined by recursion relations using the leading soft behaviour alone, for which we have verified the statements with many non-trivial examples. This can be viewed as supporting evidence for the equivalence of scale and conformal symmetry. 

This paper is organized as follows: in section \ref{Sec:soft}, we give a review of soft theorems for spontaneous symmetry breaking, and show that the mixing of GB modes under the broken symmetry can lead to non-vanishing soft limits, in contrast to the usual Adler's zero. Explicit tests for the new soft theorems were conducted in subsection \ref{Sec:OneLoop} on the one-loop and \ref{section:instantoncheck} for the instanton effective action. In section \ref{Sec:Recur}, we consider to which extent the matrix element of the dilaton effective action is fixed via soft and factorization constraints. In section \ref{Sec:SUSY}, we consider further constraints from maximal supersymmetry. In section \ref{section:scalevsconformal}, we study scale vs conformal symmetry in the context of soft-theorems. 
We conclude in section \ref{section:conclusion}.

\section{Soft theorems}\label{Sec:soft}
\begin{figure}
\begin{center}
\includegraphics[scale=0.55]{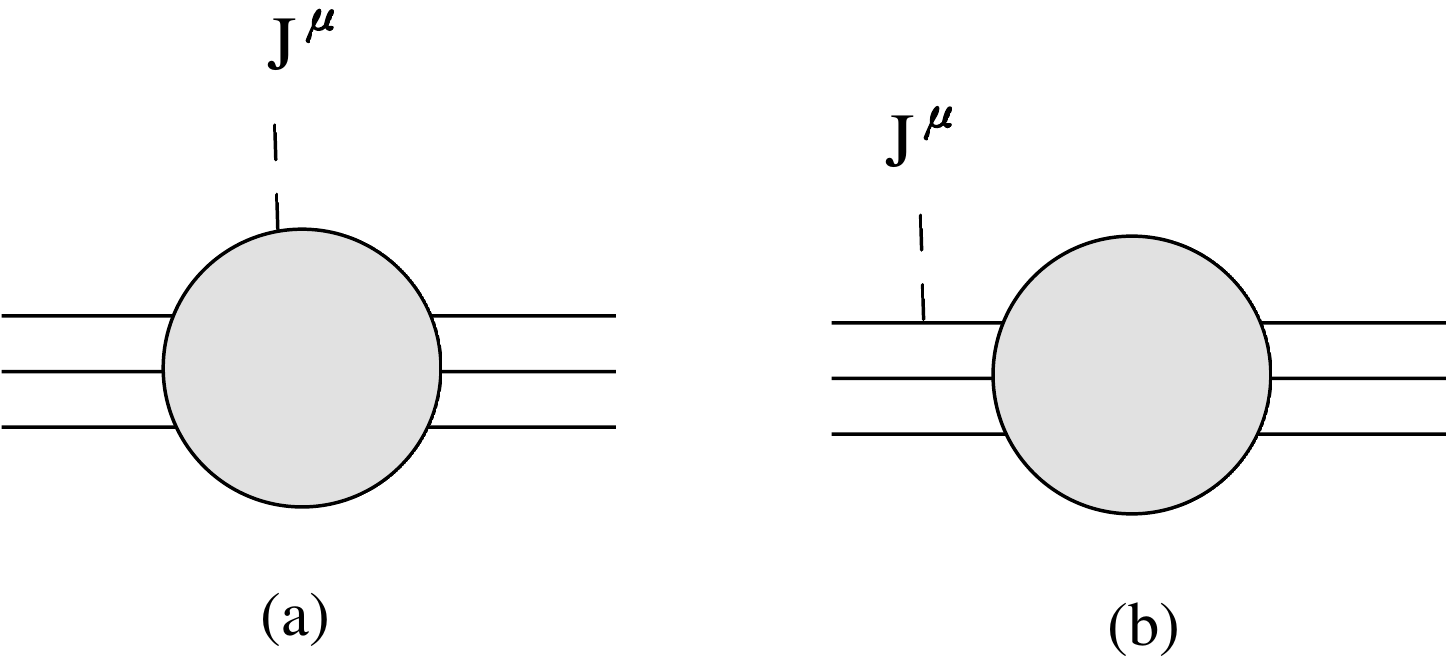}
\caption{Contributions to the soft limit}
\label{Fig1}
\end{center}
\end{figure}

Soft behaviour of amplitudes with massless particles are often dictated by Ward identities of the underlying symmetries. Here we follow the discussion in~\cite{Weinberg}, and clarify where one departs from the usual Adler's zero. Spontaneous broken symmetry implies that the current associated with the broken generators excite GBs from the vacuum:
\eq
 \langle \pi^{a}(q)|J^{b\mu}(x)|0\rangle=i f_\pi q^\mu e^{iqx} \delta^{ab}
\eqe 
where $a,b$ label the generators. Inserting the current between a set of incoming and out going asymptotic states ($\alpha,\beta$), one finds, with $q^\mu=p^\mu_\alpha-p_\beta^\mu$ 
\eq\label{soft1}
\langle \alpha| J^\mu(0)|\beta\rangle=\frac{q^\mu}{q^2}A(\pi,\alpha,\beta)+N^\mu
\eqe  
where the RHS is understood as an expansion in $q$ and we've separated out the pole term for the emission of a GB, which corresponds to fig.\ref{Fig1}(a), and $A$ is the transition amplitude.

Contracting $q_\mu$ on both sides of eq.(\ref{soft1}), the LHS vanishes since the current is conserved:
\eq
0=\langle \alpha| \partial_\mu J^\mu(x)|\beta\rangle=\langle \alpha| \partial_\mu e^{iqx}J^\mu(0)|\beta\rangle=e^{iqx}q_\mu\langle \alpha| J^\mu(0)|\beta\rangle\,.
\eqe 
This implies that 
\eq\label{Soft2}
A(\pi,\alpha,\beta)=-q^\mu N_\mu\,.
\eqe
Thus in the limit where $q\rightarrow0$, the soft limit of the amplitudes involving a GB would vanish unless $q^\mu N_\mu$ is finite. This requires non-vanishing contributions from diagrams associated with fig.\ref{Fig1}(b). Note that for the latter to yield non-trivial contribution, there must be more than one massless state in the spectrum that is charged under the current, and thus form the necessary three-point vertex.\footnote{A vector current cannot couple to two identical particles.} In other words, the broken symmetry must transform a physical state to another.

The explicit form of $q^\mu N_\mu$  can be directly read off from the Ward identity:
\eq
\partial_\mu \langle J^\mu(x) \phi(x_1)\cdots\phi(x_{n{-}1})\rangle=-\sum_{i=1}^{n{-}1} \delta(x-x_i)\langle \phi(x_1) \cdots\delta\phi(x_i)\cdots\phi(x_{n{-}1})\rangle\,.
\eqe
Fourier transform on both sides leads to  
\eq
-q_\mu\langle \tilde{J}^\mu(q) \tilde\phi(p_1)\cdots\tilde\phi(p_{n{-}1})\rangle=-\sum_{i=1}^{n{-}1}\langle \tilde\phi(p_1) \cdots\delta\tilde\phi(p_i+q)\cdots\tilde\phi(p_{n{-}1})\rangle\,,
\eqe
where $\tilde{\phi}$ represents Fourier transformed field. We now perform LSZ reduction on legs $1,\cdots,n{-}1$ on both sides by multiplying $\prod_{i} p_i^2$ and taking the momenta on-shell. The RHS vanishes for generic $q$, due to one uncanceled inverse propagator from the reduction. Taking the limit $q\rightarrow0$, the RHS develops the requisite inverse propagator if $\delta \tilde\phi$ yields a physical state in the spectrum. At the same time, the LHS is simply the amplitude with one soft GB.  
Thus we see that if $\delta \tilde\phi$ does not correspond to another particle in the spectrum, then the RHS will not survive the LSZ reduction and hence vanishes. This is the Adler's zero for soft pion emission~\cite{Adler:1964um}. Indeed in these classical examples, the Goldstone bosons transforms non-linearly under the broken symmetry, and hence its infinitesimal transformation (a shift) does not yield a particle in the spectrum. On the other hand, if $\delta \phi$ does produce a particle in the spectrum then the RHS is non-zero, and is given by the sum of Fourier transformed amplitude with the $i$-th field transformed under the generator of the broken generator. This would be $q\cdot N$.

For broken conformal symmetry, one has the latter case. The broken dilatation symmetry constrains the leading term whilst the conformal boost generators constrain the sub-leading term in the soft momentum expansion. Thus amplitudes with single soft dilaton ($\varphi$) satisfy the following universal soft theorem~\cite{Boels:2015pta, DiVecchia:2015jaq}:
\bea \label{eq:softlimitCDBI}
v A_n{\big |}_{p_n \rightarrow 0} =  
\left(  \mathcal{S}^{(0)}_n + \mathcal{S}^{(1)}_n \right) A_{n{-}1}+ \mathcal{O}(p_n^2) \, ,
\eea
where the superscript indicates the degree in $p_n$ and $v$ is the vacuum expectation value of the dilaton field. The explicit form of $\mathcal{S}^{(0)}_n, \mathcal{S}^{(1)}_n$ are given by\footnote{Note that one should replace $p_{n{-}1}$ in the $n{-}1$ point amplitude by its solution to the momentum conservation $\overline{p}_{n-1} = - (\sum^{n-2}_{i=1} p_i)$.}
\bea\label{eq:softnomassS1}
\mathcal{S}^{(0)}_n &=&- \sum^{n-1}_{i=1}
 \left( p_i {\cdot} {\partial \over \partial p_i } + {D- 2 \over 2} \right) + D \, ,\cr
\mathcal{S}^{(1)}_n  &=& -  p^{\mu}_n \sum^{n-1}_{i=1} 
\left[   p^{\nu}_i { \partial^2 \over \partial {p_i^{\nu}} \partial {p_i^{\mu}} }
-
{\frac{p_{i\mu}}{2}}{ \partial^2 \over \partial {{p_i}_{\nu}} \partial {p_i^{\nu}} } 
+ {D-2 \over 2} {\partial \over \partial {p^{\mu}_i} } \right] \, .
\eea
where $D$ is the space-time dimension.

For spontaneously broken superconformal theories, the set of massless scalars comprise the dilaton as well as the GB's for the spontaneous  breaking of R-symmetry. If the dilaton is identified with one of the scalars that transforms non-trivially under the broken R-symmetry generator, following the above discussion the soft limit of the R-symmetry GB is non-vanishing. For instance, in ${\cal N} =4$ SYM, the scalars form a $\textbf{6}$ of SO(6), any one of the scalars taking a vev (say $\phi^6$) breaks R-symmetry down to SO(5), with 5 GB's associated with the broken rotation generators $R^{6I}$ with $I=1,\cdots,5$. Under this broken generator, the GB's $\phi^I$ is rotated into $\phi^6\equiv\varphi$, while  $\varphi$ is rotated into $\phi^I$ with a relative minus sign due to the antisymmetry of $R^{6I}$. Thus the soft limit of R-symmetry GB's are given as:
\bea\label{eq:softnomassS2} 
v \, A_n(\phi_1,{\cdots}, \phi_n^I){\big |}_{p_n\rightarrow 0}=  
\sum_i A_{n{-}1}({\cdots},\delta_I\phi_i,{\cdots}) + \mathcal{O}(p_n^1) \, ,
\eea
where $\phi_i$ represents either a dilaton $\varphi$ or $\phi^I$, with $\delta^I\varphi=\phi^{I}$  and $\delta_I\phi^J=-\delta_I{}^J \varphi$. In the following subsections we will verify the soft theorems by explicitly computations of scattering amplitudes one-loop and one-instaton effective action of $\mathcal{N}=4$ SYM in the Coulomb branch. 

We should add a comment at this point. In $\mathcal{N}=4$ SYM one can define a different dilaton $\hat\varphi = \sqrt{\sum_I \phi_I^2}$ that represents the radial direction in holographic contexts and coincides with the above $\varphi = \phi_6$ (up to a sign) if the other GB's are set to zero. Moreover, the orthogonal `angular' directions of $S^5 =$ SO$(6)$/SO$(5)$ would behave as {\it bona fide} pions and satisfy Adler's theorem, since they would transform non-linearly into one another and would not mix with the radial dilaton, that is a singlet of SO$(6)$. While this is not particularly useful in the $\mathcal{N}=4$ SYM context, since it would spoil the beautiful symmetry among the various scalars, for SCFT's with lower supersymmetry, such as theories holographically dual to D3-branes at Calabi-Yau singularities (CY cones), the reduced R-symmetry would not allow such a `linear' representation of the dilaton and pions as above but only the standard non-linear one, whereby the dilaton is an R-symmetry singlet (radial direction) and the pions are the angular directions of the Sasaki-Einstein base of the CY cone.

\subsection{The one-loop verification}\label{Sec:OneLoop}
As discussed in the introduction, soft theorems hold both in the presence of the massive states and in the low energy limit where the massive states are integrated away. To verify this, we construct the one-loop effective action of $\mathcal{N}=4$ {S}YM on the Coulomb branch.\footnote{The one-loop effective action has been constructed in the constant field strength limit~\cite{Fradkin:1982kf,Chepelev:1997av}. Here we consider terms involving derivatives.} Integrands for the Coulomb branch theory can be obtained by compactifying higher-dimensional {S}YM theory, with the extra components of momenta identified with mass induced by scalar vev $v$\footnote{Obtaining spontaneously-broken SYM via a dimensional compactification was recently also studied in~\cite{Chiodaroli:2015rdg}}. We rely both on the $D=10$ {S}YM integrand constructed in~\cite{Mafra:2014gja} as well as on six-dimensional generalized unitarity methods for $(1,1)$ SYM~\cite{Dennen:2009vk, Brandhuber:2010mm} as a cross-check. At four and five points, the one-loop amplitudes of $\mathcal{N}=4$ {S}YM on the Coulomb branch are relatively simple, and have been obtained in~\cite{Chen:2015hpa},\footnote{Hereon we exploit the spinor-helicity formalism, whereby $p^{\alpha \dot{\alpha}}_i = \lambda^{\alpha}_i \tilde{\lambda}^{\dot{\alpha}}_i \,,$
and scalar products read 
\eq
\lambda_i^\alpha\lambda_j^\beta\epsilon_{\alpha\beta}=\langle ij\rangle \,, \quad \tilde\lambda_{i\dot\alpha}\tilde\lambda_{j\dot\beta}\epsilon^{\dot\alpha\dot\beta}=[ij] \,, \quad 
s_{ij}=\langle ij\rangle[ji] \, .  
\eqe }
\bea
\mathcal{A}_4  &=& {g^4 N} \, \delta^{8}(Q) {[12]^2 \over \langle 34\rangle^2} \times \sum_{S_4/Z_4} I_4(1,2,3,4; m)  \,, 
\cr
\mathcal{A}_5  &=&v {g^4 N} \, \delta^{8}(Q) {m^{(1)}_{1,2,3} m^{(2)}_{1,2,3} +
m^{(3)}_{1,2,3} m^{(4)}_{1,2,3} \over \langle 45\rangle^2} 
\times \sum_{S_5/Z_5} I_5(1,2,3,4,5; m)  \, ,
\eea
with the super charge $Q^{\alpha A}= \sum_i \lambda^{\alpha}_i \eta^A_i$. Notice that the prefactors containing fermionic $\eta$'s in both four and five points are permutation symmetric. The integrals $I_4(1,2,3,4; m)$ and $I_5(1,2,3,4,5; m)$ are scalar one-loop box and pentagon integrals with massive propagators and we sum over non-cyclic permutations, and
\bea
m^{(A)}_{i,j,k} =  [i \, j] \eta^A_k + [j \, k] \eta^A_i+ [k \, i] \eta^A_j \, .
\eea
In the above formulae the breaking of SU(4) to Sp(4) is manifest in the choice of R-symmetry indices in $m^{(A)}_{1,2,3}$, which correspond to taking the anti-symmetric $4\times 4$ Sp(4) metric to be $\Omega^{12}=-\Omega^{21}=\Omega^{34}=-\Omega^{43}=1$. In this notation, the dilaton $\Omega_{AB} \phi^{AB}$ represents fluctuations around the vev $v=m/g =\Omega_{AB} v^{AB}$. With this choice the dilaton is $\varphi = \phi^{12} + \phi^{34}$ and the other five real scalars corresponding to the pions of R-symmetry breaking are 
\bea \label{eq:Rpions}
\{\phi^{1}, \phi^{2}, \phi^{3}, \phi^{4}, \phi^{5}\} = \{i(\phi^{12}{-}\phi^{34}), \phi^{13}{+}\phi^{24}, 
i(\phi^{13}{-}\phi^{24}),\phi^{14}{+}\phi^{23}, i(\phi^{14}{-}\phi^{23})\} \, .
\eea

One can straightforwardly verify that five and four-point amplitudes do satisfy the soft theorems. Six-point amplitudes are more involved, we utilize the integrand of 10D YM obtained in~\cite{Mafra:2014gja} (especially equation $(5.10)$ in the reference) and campactify to 4D. In particular, to distinguish the dilaton from other five scalars, we set $\ell \cdot e_i =m$ if $e_i$ is dilaton and $\ell \cdot e_i =0$ if $e_i$ is one of the R-symmetry pions, here $\ell$ denotes the loop momentum and $e_i$ is the 10D polarization vector which becomes a scalar after compactification. We computed six-point amplitudes up to the order $s^5$ from the integrands by performing the integrals in the large-mass expansion, and checked the six-point amplitudes also obey the soft theorems. We have done the same computation by obtaining the corresponding integrand for $(1,1)$ SYM using the generalized unitary cuts. Some of the results will be summarized in what follows in the form of the effective action.

Although the SU(4) R-symmetry is broken down to Sp(4) on the Coulomb branch, the effective action can be conveniently decomposed into SU(4) singlet and non-singlet sectors. The one-loop effective action up to six field strengths reads
\bea\label{OneL}
\mathcal{L}^{singlet}_{\rm 1-loop} = {g^4 N \over  32  m^4\pi^2}\left(
\mathcal{O}_{F^4} + 
{ \mathcal{O}_{D^4F^4} \over 2^{3}{\times}15  m^4} 
 -{2 \mathcal{O}_{D^2F^6}   \over 15  m^{6}} +{  \mathcal{O}_{D^4F^6} \over 2^3{\times}21  m^{8}} 
- {\mathcal{O}_{D^6 F^6}  \over 2{\times}15^2  m^{10}}  + \mathcal{O}(m^{-12})\right) \cr
\eea
where $\mathcal{O}_{D^mF^n}$ represents super-local operators that contain $D^mF^n$. In the Coulomb branch $D =\partial$. Including an overall $\delta^8(Q)$, the explicit form of the superfunctions reads
\eqa
\nonumber&&\mathcal{O}_{F^4}:\:\delta^8(Q)\frac{[12]^2}{\langle34\rangle^2},\quad \mathcal{O}_{D^4F^4}:\:\delta^8(Q)\frac{[12]^2}{2\langle34\rangle^2}(\sum_{i<j}s^2_{ij})\,,\quad\mathcal{O}_{D^2F^6}:\:\frac{-\delta^8(Q)}{8}\sum_{S_6 /S_3{\times}S_3}\Xi^2_{123}\Xi^2_{456}\\
 \nonumber&&\mathcal{O}_{D^4F^6}:\:\delta^8(Q)\sum_{S_6/Z_6}\prod_i[ii{+}1],\quad\mathcal{O}_{D^6F^6}:\:\delta^8(Q)\sum_{S_6/Z_6}(\prod_i[ii+1])s_{24}\,.
\eqae
The Grassmann odd parameters $\eta^A$ appear in the super-polynomials
\eqa
\Xi^2_{123}\Xi^2_{456}=\frac{\epsilon^{ABCD}m^{(A)}_{123}m^{(B)}_{123}m^{(C)}_{456}m^{(D)}_{456}}{4!} \, .
\eqae
For the non-singlet part, we will only list the results of scalar operators which are relevant for the soft theorems we will discuss momentarily. Note that since the SO(5)$\sim$Sp(4) subgroup of R-symmetry is preserved, the $\phi^I$ pion fields must come with even multiplicity. In the following we list the result of one-loop effective action with mixed dilaton and pions,
\bea
&&\mathcal{L}^{\rm Sp(4)}_{\rm 1-loop} ={g^4 N \over 4\pi^2m^4} \left[ {{\partial}^4 \varphi^4  \over 4 } 
+{{\partial}^8 \varphi^4 \over 2^{10} \times  15 m^4} + 
{{\partial}^{10} \varphi^4 \over 2^{5} \times 3^2 \times 35 m^6}
+  {{\partial}^{12} \varphi^4 \over 2^{13} \times 3^3 \times 35  m^{8}}  -{{\partial}^4 \varphi^5 \over m^2}   
   \right.
 \cr
&&\qquad   -{{\partial}^8 \varphi^5 \over 2^7 \times 135 m^{6}} 
-{5\, {\partial}^{10} \varphi^5 \over 2^6 \times 3^4 m^{8}}
 - {{\partial}^{12} \varphi^5 \over 2^6 \times 3^5 \times 35  m^{10}}
 +{5 {\partial}^4 \varphi^6 \over  m^4} +{{\partial}^8 \varphi^6 \over 120  m^{8}} 
+{5\,{\partial}^{10} \varphi^6 \over 2^{8} 3^5   m^{10}} 
 \cr 
&& \qquad \left.  +{{\partial}^{12} \varphi^6  \over 2^{9} 3^2  m^{12}}
+ {{\partial}^4 \varphi^2 \phi^2 \over 2}  - {5  {\partial}^4 \varphi^2 \phi^4\over m^2}  + {{\partial}^4 \varphi^4 \phi^2 \over m^2} \right] + \ldots
\eea
where the on-shell matrix elements corresponding to the higher-dimensional operators are given by
\bea
{\partial}^4 \varphi^m &:\:& \sum_{ i<j } s^2_{ij} \, , \quad
{\partial}^8 \varphi^4 :\: \big( s^2_{12} + \mathcal{P}_4 \big)^2 \, ,  \quad
{\partial}^{10} \varphi^4 :\: \big( s^5_{12} + \mathcal{P}_4 \big) \, ,  \quad 
{\partial}^{12} \varphi^4 :\: \big( s^2_{12} + \mathcal{P}_4 \big)^3 \, ,  \cr
{\partial}^8 \varphi^5 &:\:& \big(s^2_{12}+ \mathcal{P}_5  \big)^2 \, , \quad
{\partial}^{10} \varphi^5 :\: {a^{(5)}_1 \over 5} + {3 \, a^{(5)}_2 \over 7} \, , \quad
{\partial}^{12} \varphi^5 :\: {a^{(6)}_1 \over 96} + a^{(6)}_2 \, , \cr
{\partial}^8 \varphi^6 &:\:& -{b^{(4)}_{1} \over 6} + {5\,b^{(4)}_{2} \over 768} 
- {3 \, b^{(4)}_{3} \over 2} + {b^{(4)}_{4}\over 36}  \, , \cr
{\partial}^{10} \varphi^6 &:\:&
{114 \over 35} b^{(5)}_{1} + {60 \over 7}  b^{(5)}_{2} 
-{48\,b^{(5)}_{3} \over 7 } 
+ {108 \over 7} b^{(5)}_{4} 
+{36 \over 35}  b^{(5)}_{5} \, , \cr
\cr
{\partial}^{12} \varphi^6 &:\:&  {433 \over 1350 } b^{(6)}_{1}- {58 \over 2025}  b^{(6)}_{2}
+ {20 \over 9}  b^{(6)}_{3}+ {117 \over 35} b^{(6)}_{4} - {184 \over 945}  b^{(6)}_{5} , \cr
\cr
&&-{74 \over 45 } b^{(6)}_{6}+ {334 \over 315}  b^{(6)}_{7}
+{73 \over 35} b^{(6)}_{8}- {64 \over 63} b^{(6)}_{9} + {104 \over 105}  b^{(6)}_{10}  \cr
{\partial}^4 \varphi^2 \phi^2 &:\:& s_{12}^2 -s_{13}^2- s^2_{23} \, , \quad
{\partial}^4 \varphi^2 \phi^4 :\: b^{(2)}_{1,S_2 \times S_4} -  b^{(2)}_{2,S_2 \times S_4} + b^{(2)}_{3,S_2 \times S_4} - {8 \over 5} b^{(2)}_{4,S_2 \times S_4} \, ,
\cr
{\partial}^4 \varphi^4 \phi^2 &:\:& b^{(2)}_{1,S_2 \times S_4} -  b^{(2)}_{2,S_2 \times S_4}
 + b^{(2)}_{3,S_2 \times S_4} + {8 } b^{(2)}_{4,S_2 \times S_4} 
\eea  
and the $b$'s are independent symmetric polynomials, they are given by
\bea
a^{(5)}_1 &=& s_{12}^5 + \mathcal{P}_5 \, , \quad 
a^{(5)}_2 = s_{12}^2 s_{34}^3 + \mathcal{P}_5 \, ,\quad
a^{(6)}_1 = (s_{12}^2 + \mathcal{P}_5)^3 \, , \quad 
a^{(6)}_2 = s_{12}^2 s_{34}^4 + \mathcal{P}_5 \, ,
\cr
b^{(4)}_{1} &=& s_{12}^4 + \mathcal{P}_6 \, , \quad 
b^{(4)}_{2} = (s_{12}^2 + \mathcal{P}_6)^2 \, , \quad
b^{(4)}_{3} = s_{12}^2 s_{13}^2 + \mathcal{P}_6 \, , \quad
b^{(4)}_{4} =  s_{123}^4 + \mathcal{P}_6 \, , \cr
b^{(5)}_{1} &=& s_{12}^5 + \mathcal{P}_6 \, , \quad 
b^{(5)}_{2} = s_{12}^2 s_{123}^3 + \mathcal{P}_6 \, , \cr
b^{(5)}_{3} &=& s_{12}^2 s_{23}^3 + \mathcal{P}_6 \, , \quad 
b^{(5)}_{4} =  s_{12}^2 s_{34}^2 + \mathcal{P}_6 \, , \quad 
b^{(5)}_{5} =  s_{123}^5 + \mathcal{P}_6 
\label{basis}\\
b^{(6)}_{1} &=& s_{12}^6 + \mathcal{P}_6 \, , \quad 
b^{(6)}_{2} = s_{123}^6 + \mathcal{P}_6 \, , \quad 
b^{(6)}_{3} = s_{12}^4 s_{13}^2 + \mathcal{P}_6 \, , \cr
b^{(6)}_{4}&=&  s_{12}^4 s_{34}^2 + \mathcal{P}_6 \, , \quad
b^{(6)}_{5} = s_{12}^3 s_{13}^3 + \mathcal{P}_6,\quad
b^{(6)}_{6} = s_{12}^3 s_{34}^3 + \mathcal{P}_6 \, , \cr 
b^{(6)}_{7} &=& s_{12}^2 s_{123}^4 + \mathcal{P}_6 \, , \quad 
b^{(6)}_{8} = s_{14}^2 s_{123}^4 + \mathcal{P}_6 \, , \quad 
b^{(6)}_{9} =  s_{14}^4 s_{123}^2 + \mathcal{P}_6 \, , \cr 
b^{(6)}_{10} &=& s_{123}^2 s_{124}^2 s_{135}^2 + \mathcal{P}_6,\quad
b^{(2)}_{1,S_2 \times S_4} = s_{12}^2+ \mathcal{P}_{\{12|3456\}}  \, , \quad 
b^{(2)}_{2,S_2 \times S_4} = s_{13}^2 + \mathcal{P}_{\{12|3456\}} \, , \cr
\!\!\!\!\!\!\!\!\! b^{(2)}_{3,S_2  \times S_4}  &=& s_{34}^2 + \mathcal{P}_{\{12|3456\}} \, , \quad 
b^{(2)}_{4,S_2 \times S_4} =  s_{12}s_{13} + \mathcal{P}_{\{12|3456\}} \, , \nonumber
\nonumber
\eea
here $\mathcal{P}_n$ denotes summing over permutations of $n$ elements, while $\mathcal{P}_{\{n|m\}}$ denotes summing over permutations of $n$ and $m$ elements.

\subsection{Non-perturbative checks} \label{section:instantoncheck}
Relying on (unoriented) open strings and D-brane instantons\footnote{See e.g. \cite{Bianchi:2007ft, Bianchi:2009ij, Bianchi:2012ud} for recent reviews.}, the one-instanton corrections to the effective action of $\mathcal{N}=4$ SYM in the Coulomb branch have been computed in~\cite{Bianchi:2015cta}.
For Sp(2N) the integration over (super)moduli space can be performed, and the resulting effective action can be written in a very compact and elegant form 
\bea   
S^{1-inst}_{\rm eff} =  c' \, {g^4\over \pi^6} \, e^{2\pi {\rm i} \tau } \, \int{  { d^4 x\, d^8 \theta\, \sqrt{{\rm det}_{4N} \, 2\bar\Phi_{Au,Bv} }  \over 
\sqrt{{\rm det}_{2N} \left( \Phi^{AB}\bar\Phi_{AB}  +{1\over g} {\bar{\mathcal{F}} }  +\frac{1}{\sqrt{2} g} \bar{\Lambda}_{A} (\Phi^{-1})^{AB} \bar{\Lambda}_{B}  \right)_{\dot\alpha u, \dot\beta v} }  } }\, ,
\label{seff0}
\eea
where $\tau = (\vartheta/2\pi) + (4\pi i/g^2)$ is the complexified coupling and the ${\cal N} =4$ on-shell superfields can be expanded in terms of the component fields $\{\phi^{AB}, \lambda^A_\alpha, F^-_{\alpha\beta}\}$ and their conjugate according to
\begin{align}
\bar\Phi_{AB}&=\bar\phi_{AB}+\varepsilon_{ABCD}\theta^{C\alpha}\lambda^D_{\alpha}+\frac{1}{2}\varepsilon_{ABCD}\,\theta^{C\alpha} F^{-}_{\alpha\beta}\theta^{D\beta}\\
\bar\Lambda_{\dot\alpha A}&=\bar\lambda_{\dot\alpha A}+i\,\theta^{B\alpha}\de_{\alpha\dot\alpha}\bar\phi_{AB}+\frac{i}{2}\varepsilon_{ABCD}\theta^{B\beta}\theta^{C\gamma}\de_{\{ \beta\dot\alpha} \lambda^D_{\gamma\}}+\frac{i}{6}\varepsilon_{ABCD}\theta^{B\alpha}\theta^{C\beta}\theta^{D\gamma}\de_{\{\alpha\dot\alpha}F^-_{\beta\gamma\}}\\
\bar{\mathcal{F}}_{\dot\alpha\dot\beta}&=F^+_{\dot\alpha\dot\beta}-i\,\theta^{A\alpha}\de_{\alpha\{ \dot\alpha} \bar\lambda_{A\dot\beta\}}+\frac{1}{2}\theta^{A\alpha}\theta^{B\beta}\de_{\alpha\dot\alpha}\de_{\beta\dot\beta}\bar\phi_{AB}+\frac{1}{6}\varepsilon_{ABCD}\theta^{A\alpha}\theta^{B\beta}\theta^{C\gamma} \de_{\alpha\dot\alpha}\de_{\beta\dot\beta}\lambda^D_\gamma\nonumber\\
&-\frac{1}{24}\varepsilon_{ABCD}\theta^{A\alpha}\theta^{B\beta}\theta^{C\gamma}\theta^{D\delta}\de_{\alpha\dot\alpha}\de_{\beta\dot\beta}F^-_{\gamma\delta}.
\end{align}
For the study of soft-dilaton and soft-pion theorems, we will turn on just the scalar fields so that
\be
\bar\Phi_{AB}=\bar\phi_{AB} \, , \quad \bar\Lambda_{A\dot\alpha}=i\,\theta^{B\alpha}\de_{\alpha\dot\alpha}\bar\phi_{AB}\, , \quad \bar{\mathcal{F}}_{\dot\alpha\dot\beta}=\frac{1}{2}\theta^{A\alpha}\theta^{B\beta}\de_{\alpha\dot\alpha}\de_{\beta\dot\beta}\bar\phi_{AB} \, ,
\ee
and\footnote{$\bar\phi_{AB} = {1\over 2} \varepsilon_{ABCD}\phi^{CD}$, $\phi^2 = \sum_i \phi_i^2 = {1\over 4}\bar\phi_{AB}\phi^{AB}$} 
$(\bar\Phi^{-1})^{AB}={\phi^{AB} }/{ \phi^2}$.
As a result the one-instanton effective action drastically simplifies and takes the following form
\be
S^{1-inst}_{\rm eff} = c' {g^4 \over \pi^6} e^{2\pi i \tau} \int {d}^4x\, {d}^8\theta\,\frac{1}{1- H_{\dot\alpha\dot\beta}H^{\dot\alpha\dot\beta}}= c' {g^4 \over \pi^6} e^{2\pi i \tau} \int {d}^4x\, {d}^8\theta\, (H_{\dot\alpha\dot\beta}H^{\dot\alpha\dot\beta})^2,
\ee
where 
\be
H_{\dot\alpha\dot\beta}=\frac{1}{g \, \Phi^2}\left(\bar{\mathcal{F}}_{\dot\alpha\dot\beta}+{1\over\sqrt{2}}\bar\Lambda_{A\dot\alpha}(\bar\Phi^{-1})^{AB}\bar\Lambda_{B\dot\beta}\right) \, .
\ee
In the last step we have expanded the denominator and only kept the term which is non-vanishing after Grassman integration if one takes into account that the super-field $H_{\dot\alpha\dot\beta}$ becomes
\be
H_{\dot\alpha\dot\beta}=\frac{1}{4 \, g \, \phi^2}\left(\frac{1}{2} \de_{\alpha\dot\alpha}\de_{\beta\dot\beta}\phi_{AB} -\frac{\phi^{CD}\de_{\alpha\dot\alpha}\phi_{A C}\de_{\beta\dot\beta}\phi_{DB}}{\phi^2} \right) \theta^{A\alpha}\theta^{B\beta} = K_{\alpha\dot\alpha\beta\dot\beta, AB} \theta^{A\alpha}\theta^{B\beta}\, ,
\ee
when only scalars are turned on as in the case of interest here.
Switching to 4-vector indices $K_{\alpha\dot\alpha\beta\dot\beta, AB}$ may be decomposed into a symmetric traceless tensor $S^{(\mu\nu)}_{[AB]}$ in the ${\bf 6}$ of SU(4)$\sim$SO(6) and an anti-symmetric
tensor $B^{[\mu\nu]}_{(AB)}$ in the ${\bf 10^*}$ of SU(4)$\sim$SO(6). For instance, for pure dilaton sector only symmetric tensor $S^{(\mu\nu)}_{[AB]}$ contributes, and after performing the fermionic integration the action is given as,
\be
\mathcal{S}_{\rm dilaton}=\int d^4x\,\left[ (S_{\mu\nu}S^{\mu\nu})^2-S_{\mu\nu}S^{\nu\rho}S_{\rho\sigma}S^{\sigma\mu} \right].
\ee
where
\be
S_{\mu\nu}=\frac{\de_\mu\de_\nu\varphi}{\varphi^2}-2\frac{\de_\mu\varphi\de_\nu\varphi}{\varphi^3}-\frac{1}{4}\eta_{\mu\nu}\frac{\de^2\varphi}{\varphi^2}+\frac{1}{2}\eta_{\mu\nu}\frac{\de_\rho\varphi\de^\rho\varphi }{\varphi^3} \, ,
\ee
and the dilaton $\varphi$ has a non-vanishing vev $\varphi \rightarrow \varphi + v.$
With the one-instanton action at hand, we have computed amplitudes up to seven points for dilaton and pion mixed amplitudes and pure-dilaton amplitudes up to nine points. We find that they indeed satisfy all the soft theorems. Here we list a few pure dilaton amplitudes\footnote{The overall coupling dependence such as $e^{2 \pi i \tau}$ is understood.}, which are degree-four symmetric polynomials in $s_{ij}$,  
\begin{align} \label{inst_amps}
v^8 \mathcal{A}_4^{inst}&=\frac{1}{32} \left( s_{12}^2 + \mathcal{P}_4 \right)^2 \, , \quad
v^9 \mathcal{A}_5^{inst}=-\frac{1}{36} \left( s_{12}^2 + \mathcal{P}_5 \right)^2 \, ,\\ \nonumber
v^{10} \mathcal{A}_6^{inst}&=-\frac{2}{3} b_1^{(4)}  +\frac{5}{192} b_2^{(4)}
- 6 \,  b_3^{(4)} +\frac{1}{9} b_4^{(4)}  \, ,\\ \nonumber
v^{11} \mathcal{A}_7^{inst}&=4 \,b_{1,7}^{(4)}  + 40  \,b_{2,7}^{(4)}  
-\frac{5}{3}  b_{3,7}^{(4)}  
-25  \,b_{4,7}^{(4)}  \, , \\ \nonumber
v^{12} \mathcal{A}_8^{inst} &= -{809\over 144} b_{1,8}^{(4)} - {395\over 8} b_{2,8}^{(4)} +
{1339\over 576} b_{3,8}^{(4)} + {595\over 32} b_{4,8}^{(4)} + {535\over 32} b_{5,8}^{(4)}
\, , \\ \nonumber
v^{13} \mathcal{A}_9^{inst} &= {3935\over 294} b_{1,9}^{(4)} +{846\over 7} b_{2,9}^{(4)} -
{475\over 126} b_{3,9}^{(4)} - {491\over 14} b_{4,9}^{(4)} - {535\over 14} b_{5,9}^{(4)} \, , \nonumber
\end{align}
where the six-point amplitude $\mathcal{A}_6^{inst}$ is expanded in the basis given by eq.(\ref{basis}), while for the higher-point amplitudes 
\bea
b^{(4)}_{1,7} &=& s_{12}^4 + \mathcal{P}_7 \, , \quad 
b^{(4)}_{2,7} = s_{12}^2s_{23}^2 + \mathcal{P}_7 \, , \quad 
b^{(4)}_{3,7} = s_{123}^4 + \mathcal{P}_7 \, , \quad 
b^{(4)}_{4,7} = s_{123}^2s_{124}^2 + \mathcal{P}_7 \, ,\cr 
b^{(4)}_{1,8} &=&  s_{12}^4 + \mathcal{P}_8 \, , \quad
b^{(4)}_{2,8} = s_{12}^2s_{23}^2 + \mathcal{P}_8 \, , \quad 
b^{(4)}_{3,8} = s_{123}^4  + \mathcal{P}_8 \, , \quad
b^{(4)}_{4,8} = s_{123}^2s_{124}^2 + \mathcal{P}_8\, , \cr
b^{(4)}_{5,8} &=& s_{123}^2s_{145}^2  + \mathcal{P}_8\, ,\cr 
b^{(4)}_{1,9} &=&  s_{12}^4 + \mathcal{P}_9 \, , \quad
b^{(4)}_{2,9} = s_{12}^2s_{23}^2 + \mathcal{P}_9 \, , \quad 
b^{(4)}_{3,9} = s_{123}^4  + \mathcal{P}_9 \, , \quad
b^{(4)}_{4,9} = s_{123}^2s_{124}^2 + \mathcal{P}_9\, , \cr
b^{(4)}_{5,9} &=& s_{123}^2s_{145}^2  + \mathcal{P}_9.
\eea
In appendix \ref{Appendix:oneinstanton}, we have also listed the higher-dimensional vertices that generate the above amplitudes. As we mentioned we have verified that all these amplitudes indeed satisfy the soft theorems. In fact, as we will discuss in section \ref{section:N=4SUSY}, at four, five and six points, amplitudes (with both dilatons and pions) at order $s^4$ are fully fixed by $\mathcal{N}=4$ SUSY and the soft theorems. Furthermore, for the pure-dilaton amplitudes all the higher-point amplitudes at this order are fully determined by the soft-dilaton theorems from the knowledge of the five-point amplitude, as we will discuss in the next section. Thus consistency with the conformal symmetry and $\mathcal{N}=4$ SUSY (which fixes the form of the five-point amplitude), the pure-dilaton amplitudes in fact must take the unique form given in (\ref{inst_amps}), and the same holds true for higher-point ones.

\section{Constraining the effective actions by means of soft theorems}\label{Sec:Recur}
An immediate consequence of the dilaton soft theorem is its constraint on the effective action. A systematic way to explore soft constraints is the recently constructed on-shell recursion relations~\cite{Cheung:2015ota, Luo:2015tat}. On-shell recursive methods are constructed using the fact that under complex deformation of the momenta, the only allowed singularities are propagator singularities whose residues are determined by lower point data. Using the fact that S-matrix elements are analytic functions, we start with~\cite{Britto:2004ap, Britto:2005fq}:
\eq
A_n(0)= {1 \over 2 \pi i} \oint_{{\cal C}_0} dz \frac{A_n(z)}{z}\,,
\eqe
where the contour ${\cal C}_0$ encircles the origin, and $A_n(z)$ is the $n$-point amplitude with deformed momenta and $A_n(0)$ is the undeformed amplitude which we would like to compute. If $A_n(z)$ is meromorphic, via the residue theorem, we can recast the amplitude as a sum over residues at finite values in the complex plane plus the one at infinity. The poles at finite values in the complex plane are simply due to factorization and their residues are determined by lower-point amplitudes. The usefulness of the recursion then relies on whether one can avoid contributions from the point at infinity or one can determine that contribution {\it a priori}. Effective theories in general do receive contributions at infinity. In~\cite{Cheung:2015ota, Luo:2015tat}, it was shown that if it is known that the amplitude has universal behaviour in some kinematic regime, then one can instead consider
\eq\label{BCFW}
A_n(0)= {1 \over 2 \pi i}  \oint_{{\cal C}_0} dz  \frac{A_n(z)}{z F(z)} \, ,
\eqe
where $F(z)$ is a polynomial in $z$ with $F(0)=1$, and its zeroes correspond to the special kinematic configurations. At large $z$, $F(z) \sim z^d$ with some positive $d$. The function $F(z)$ introduces extra power of suppression at large $z$, allowing for non-vanishing boundary contributions from $A(z)$ of higher mass dimension. The amplitude $A(0)$ is then determined by the residues of the factorisation pole as well as the contributions from the poles in $1/F(z)$ which are given by the universal behaviour of the amplitudes.   

For theories with universal soft theorems, one deforms the amplitude by shifting each momentum as $p_i\rightarrow (1-z a_i) p_i$, such that $z \rightarrow 1/a_i$ one approaches the soft limit. This leads to the choice of $F(z)$ given in~\cite{Cheung:2015ota, Luo:2015tat}
\eq
F_n(z)=\prod_{i=1}^n(1-z a_i)^{d_i}\,,
\eqe
where $\sum_{i}a_i p_i=0$ to ensure momentum conservation for $A(z)$, and the positive integer $d_i$ depends on the soft theorem for the particle species of external leg $i$: particle $i$ enjoys universal soft theorem up order $q^{d_i-1}$ in the soft momentum ($q$) expansion. That is because otherwise $F_n(z)$ would introduce poles whose residues would be unknown. 

\begin{table}[t]
\centering 
\begin{tabular}{c|ccccccccccc}
$s^n$  $\backslash$  \# of points  &  4 ~ & 5 ~ & 6 ~ & 7 ~ & 8 ~ & ~ $\cdots$ ~
 \\ \hline
2 &  $\times$ & \checkmark & \checkmark & \checkmark & \checkmark & \checkmark 
  \\
3 &  $\times$ & \checkmark & \checkmark & \checkmark & \checkmark & \checkmark 
  \\
4 &  $\times$  & \checkmark & \checkmark & \checkmark & \checkmark & \checkmark 
  \\
5 &  \checkmark & $\times$ & \checkmark & \checkmark & \checkmark & \checkmark 
  \\
6 &  \checkmark & \checkmark & $\times$  & \checkmark & \checkmark & \checkmark 
  \\
7 &  \checkmark & \checkmark & \checkmark & $\times$  & \checkmark & \checkmark
  \\
8 &  \checkmark & \checkmark & \checkmark & \checkmark  & $\times$ & \checkmark
 \\
\vdots & $\cdots$  & $\cdots$ & $\cdots$ & $\cdots$  & $\cdots$ & $\cdots$
\end{tabular}
\caption{The table is to show that the knowledge of the $k$-point amplitude at order $s^k$ with $k \leq n$ allows one to determine all the amplitudes up to the order $s^n$. The $\times$ is to indicate the amplitudes that have to compute by other means, then all other amplitudes marked with \checkmark are completely determined by the soft theorems as well as the soft-BCFW. One should note that the soft-BCFW recursion relation can only apply to amplitudes in a $D$-dimensional theory with at least $D{+}2$ external legs.
\label{tab:summary}}
\end{table}

Let us first consider the dilaton effective action. Since the dilaton soft theorem is universal up to $\mathcal{O}(q^1)$, this implies that all the $d_i$ can be mostly taken to be $2$, and thus $F_n(z)$ behaves as $z^{2n}$ in the large $z$ limit. Simple power counting shows that for an amplitude at order $s^k$, the recursion formula is valid for $A_n$ if $n>k$. Thus the pure dilaton sector is completely determined if the $n$-point amplitude at order $s^n$ is known, as the higher-point amplitudes are uniquely determined via recursion, while lower-point amplitudes can simply be obtained through leading soft theorems by taking a soft particle away. Therefore if the $k$-point amplitude at order $s^k$ with all $k \leq n$ are given, one can determine all the amplitudes up to the order $s^n$. One should also take into account that the soft BCFW recursion relation is only applicable in $D$-dimensions for at least $D{+}2$ external legs. For instance, in $D=4$, at order $s^4$, knowing the four-point amplitude is not enough to completely fix all higher-point amplitudes. Instead the five-point amplitude is required to fully determine all amplitudes at this order. This general discussion is summarized in table \ref{tab:summary}. For general superconformal theories, one can consider mixed amplitudes with $n_1$ dilatons and $n_2$ R-symmetry Goldstone bosons. Since the R-symmetry soft theorem is only leading, the requisite bound for valid recursion is $n_1+{1\over 2} n_2 > k$ for order-$s^k$ amplitudes. 

In some special cases, all the terms marked with ``$\times$" in table \ref{tab:summary}, except the one of order $s^2$, may simply vanish. For instance, this is indeed the case if each scalar insertion carries at most one derivative, namely analogous to the ``constant field-strength approximation''. Thus at $2n$ or $(2n{+}1)$ points, amplitudes go as $s^n$ at most. Let us normalize the four-point amplitude at order $s^2$ as
\bea
A^{(2)}_4 = c^{(2)}_4 \left( s^2 + t^2 + u^2  \right) \, ,
\eea
then all the amplitudes are completely determined by the soft theorems in terms of the factor $c^{(2)}_4$, which must be non-zero and positive for a non-trivial interacting theory~\cite{Adams:2006sv}. In other words, the theory in the ``constant field-strength approximation'' is uniquely fixed by soft theorems of the (broken) conformal symmetry, which turns out to be the conformal DBI, {\it i.e.} the DBI action in the AdS background, with an appropriate choice of the overall coefficient $c^{(2)}_4$. This conclusion is in the analog of the analysis of~\cite{Cheung:2014dqa}, where the usual flat-space DBI (namely DBI action in the flat-space background) are uniquely determined by the so-called enhanced soft limits. We remark that the flat-space DBI is a special limit of the conformal DBI. The scattering amplitudes in flat-space DBI are in a subset of those in conformal DBI, in particular the highest-derivative amplitudes with even number of external legs.

\section{Constraints from supersymmetry}\label{Sec:SUSY}
Supersymmetry imposes further constraint by relating coefficients of higher dimension operators with different dimensions. More precisely, one considers matrix elements of a susy invariant local operator of a given dimension and multiplicity. If one can conclude that such an operator does not exist, then the contribution of local operators must be proportional to the one produced by factorization channels. This leads to non-renormalization conditions. Indeed recently a whole set of new non-renormalization theorems have been obtained for the effective action of supersymmetric gauge and gravity theories in diverse dimensions~\cite{Wang:2015jna, Lin:2015ixa, Cordova:2015vwa, Chen:2015hpa}. Here we will consider the consequences of combining constraints from maximal SUSY and soft theorems. 

\subsection{4D $\mathcal{N}=4$ supersymmetry} \label{section:N=4SUSY}
\subsubsection{Pure dilaton sectors}
Already for $\mathcal{N}=4$ {S}YM, it was shown that operators of the form $F^2_-F^{2\ell}_+$ is $\ell$-loop exact~\cite{Chen:2015hpa}, with the coefficient recursively determined by that of the four-point $F^2_-F^{2}_+$ operator, which was known to be one-loop exact~\cite{Dine:1997nq}. As already shown at one-loop order, generally SUSY invariant local operators for four and five points take the form:\footnote{Note four- and five-point amplitudes admit no factorization channels.}
\bea \label{eq:45pts}
\mathcal{A}_4 &=& \delta^{8}(Q) {[12]^2 \over \langle 34\rangle^2} \sum_k P^{(k)}_4(s_{ij}) \, , \\
\mathcal{A}_5 &=&v\, \delta^{8}(Q) {m^{(1)}_{1,2,3} m^{(2)}_{1,2,3} +
m^{(3)}_{1,2,3} m^{(4)}_{1,2,3} \over \langle 45\rangle^2} \sum_k P^{(k)}_5(s_{ij})\,,
\eea
where $P^{(k)}_n(s_{ij})$ represents $n$-point degree-$k$ symmetric polynomials of $s_{ij}$. As well-known, for high enough $k$, the polynomials may have diverse structures. In particular $P^{(k)}_4(s_{ij})$ starts to have two independent structures at $k=6$, and $P^{(k)}_5(s_{ij})$ has two structures at $k=4$. Maximal SUSY relates purely gluonic operators $\partial^kF^n$ to operators with scalars $\partial^{k+n} \phi^n$.
As one can see, at least for four and five points, due to maximal SUSY, a degree $k$ operator in $s$ is determined by a polynomial of degree $k{-}2$ which generally has fewer degrees of freedom. This is crucial for mixed operators with both pions and dilaton, which would otherwise not even have the full permutation symmetry. As we will see, this simple fact leads to further non-renormalization theorems: the dilaton effective action up to 10 derivatives is completely determined by two unknown coefficients of the four-point operator at $s^4$ and $s^5$.

For the four-point amplitudes of order $s^2$, the four-point result is one-loop exact, namely it does not receive higher-loop and non-perturbative corrections~\cite{Dine:1997nq}: 
\bea \label{eq:4pts2term}
P^{(0)}_4(s_{ij}) = c^{(0)}_4(g, N) \times 1\,,
\eea
where the one-loop exact coefficient $c^{(0)}_4(g, N) = {g^4 N\over 32\pi^2 m^4}.$
By knowing the four-point amplitude of order $s^2$, the dilaton soft theorems allow us to determine all higher-point amplitudes at this order. Since there are no factorization contributions in the recursion, all coefficients are determined by the four-point amplitude and hence one-loop exact. One can easily see that these higher-point amplitudes are identical to that derived from DBI action in AdS background, that we dubbed conformal DBI earlier on. 

For amplitudes at $s^3$, the four and five-point matrix element is simply zero due to the fact that $P^{(1)}_5(s_{ij})=0$ from momentum conservation. Thus the first non-trivial amplitude starts at six-point which is constructible via soft-dilaton recursion. The six-point amplitude receives contributions from local operator ${\partial}^6\phi^6$ as well as from factorization, which can be parametrized as
\bea
A^{(3)}_6 &=&
 a_1 (s_{12}^3 + \mathcal{P}_6) + a_2 (s_{123}^3 + \mathcal{P}_6) 
\cr
&+& 
\left({g^4 N\over 8\pi^2 m^4} \right)^2 \left( 
(s_{12}^2+s_{13}^2 + s_{23}^2) {1 \over s_{123}} (s_{45}^2+s_{46}^2 + s_{56}^2) + \mathcal{P}_6 \right) \,,
\eea
where we have used the result of (\ref{eq:4pts2term}). The soft theorems then fix
\bea
a_1 =0 \, , \quad a_2 = - \left({g^4 N\over 8\pi^2 m^4} \right)^2 \,.
\eea
We see that the soft theorems fix the coefficient of the local 6-point operator to be the square of that of the 4-point operator of order $s^2$. Since the latter is one-loop exact, the six-derivative operator ${\partial}^6\phi^6$ as well as the amplitude $A^{(3)}_6$ are two-loop exact. The same analysis applies to amplitudes beyond six points, and the recursion implies that order $s^3$ amplitudes for arbitrary multiplicity are two-loop exact. In terms of higher-dimensional operators, the soft theorems fix all the four and six-derivative operators (${\partial}^4 \phi^n$ and ${\partial}^6 \phi^n$) completely, and they are in fact identical to the conformal DBI. 

At order $s^4$, an $n$-point amplitude receives contributions from factorization diagrams at order $s^3$ and $s^2$, as well as the contribution from the local operator ${\partial}^8 \phi^n$. As the factorization contributions are identical to those of conformal DBI, it is convenient to separate the contribution from the local operator ${\partial}^8 \phi^n$ into two parts: DBI and non-DBI.\footnote{Such separation was also used for the operator $F^8 \sim F^2_-F^6_+ + F^2_+F^6_- + F^4_-F^4_+$ in~\cite{Buchbinder:2001ui,Chen:2015hpa}, where the first two ``MHV" and ``anti-MHV" operators are three-loop exact, and coincide with DBI.} In this way, the local DBI part combining with factorization channels reproduces the amplitudes generated from conformal DBI, which is three-loop exact at order $s^4$. This separates the amplitude into two independent solutions to the soft equations. The remaining non-DBI part consists of degree-$4$ symmetric polynomials in $s_{ij}$. Again since at this order the amplitude is recursively constructible beyond four points, the non-DBI contribution is completely determined by the coefficient of the four-point operator, which is unique at this order,\footnote{$c^{(2)}_{4}(g,N)$ as well as $c^{(3)}_4(g,N)$ that will appear later at order $s^5$ have been computed at one and two-loop orders in~\cite{Bianchi:2015cta}.} 
\bea
P^{(2)}_4(s_{ij}) =  c^{(2)}_{4}(g,N) \left( s_{12}^2 + \mathcal{P}_4 \right) \, .
\eea
Note that non-DBI contributions will be identical to that of the one-loop effective action (since there is no factorization at one loop) up to an overall normalization, namely $c^{(2)}_{4}(g,N)$ at one-loop order, thus we denote this part as $\mathcal{L}^{\ell{=}1}_{{\partial}^8\phi^n}$. 

In summary, up to order $s^4$, the dilaton effective action is constrained by $\mathcal{N}=4$ supersymmetry as well as the soft theorems to take the form 
\bea \label{eq:1-loopexact}
\sum_{k \leq 8} \mathcal{L}_{{\partial}^k\phi^n} = \delta_{k, 8}\, c^{(2)}_{4}(g,N)\mathcal{L}^{\ell{=}1}_{{\partial}^8\phi^n} 
+ \sum_{k \leq 8} \mathcal{L}^{\rm DBI}_{{\partial}^k\phi^n} \, ,
\eea
namely when $k<8$, the on-shell action is identical to conformal DBI, and at order $k=8$ the all-loop and non-perturbative action is fully determined by a single coefficient $c^{(2)}_{4}(g,N)$ of the four-point amplitude at this order. 

For the amplitudes at order $s^5$ there is again a single polynomial both at four and five points, namely
\bea
P^{(3)}_4(s_{ij}) = c^{(3)}_4(g,N) \times (s_{12}^3 +  \mathcal{P}_4 ) \,, \quad 
P^{(3)}_5(s_{ij}) = c^{(3)}_5(g,N) \times ( s_{12}^3 +  \mathcal{P}_5)\, .
\eea 
First of all, the soft theorems requires $c^{(3)}_5(g,N)=-2c^{(3)}_4(g,N)$. From soft-BCFW recursion relations, at order $s^5$ knowing the five-point amplitude allows us to fix the amplitudes of arbitrary multiplicity. At this order, the factorization contributions come from amplitudes of order $s^2$, $s^3$ as well as $s^4$. As we have argued the amplitudes of order $s^2$, $s^3$ are one and two-loop exact and coincide with conformal DBI, while order-$s^4$ amplitudes we separate into DBI and non-DBI parts. So it is again convenient to separate a DBI part from the ten-derivative operator $\partial^{10} \phi^n$, such that it combines with factorization diagrams from the amplitudes of order $s^2$, $s^3$ as well as DBI part of the order-$s^4$ amplitudes, and generates the corresponding amplitude of conformal DBI at this order, which is four-loop exact. 

Let us now consider the remaining contributions, which contain the factorization terms of the non-DBI part of the order-$s^4$ amplitudes with the amplitudes at order $s^2$, as well as non-DBI part of the local operator $\partial^{10} \phi^n$. Due to the fact that amplitudes of order $s^2$ and of non-DBI part of the order-$s^4$ are both in the one-loop form, (\ref{eq:1-loopexact}), and clearly they produce the factorization parts that are in the same form as those of two-loop amplitudes. Thus these factorizations can be conveniently combined with a piece from the non-DBI part of $\partial^{10} \phi^n$ to produce the amplitudes as two-loop ones (again up to an overall factor), which we denote as $\mathcal{L}^{\ell{=}2}_{{\partial}^{10}\phi^n}$. The above analysis leads to the following compact representation for the complete $\partial^{10}$ effective action, 
\bea
\mathcal{L}_{{\partial}^{10}\phi^n} = c^{(3)}_{4}(g,N)\mathcal{L}^{\ell{=}1}_{{\partial}^{10}\phi^n} + c^{(0)}_{4}(g,N) \times c^{(2)}_{4}(g,N)\mathcal{L}^{\ell{=}2}_{{\partial}^{10}\phi^n} 
+\mathcal{L}^{\rm DBI}_{{\partial}^{10}\phi^n} \, ,
\eea
again the kinematics dependences of $\mathcal{L}^{\ell{=}1}_{{\partial}^{10}\phi^n}$ and $\mathcal{L}^{\ell{=}2}_{{\partial}^{10}\phi^n}$ are completely fixed, and identical to those of the effective action at one and two loops, respectively. 

Beyond order $s^5$, the symmetric polynomials $P_4(s_{ij})$ and $P_5(s_{ij})$ can in general be expressed in terms of several independent structures, which may differ at different loop orders and instanton levels. Furthermore, in order to apply the soft-BCFW recursion relations one eventually requires the knowledge of amplitudes beyond four points. For instance at order $s^6$, $P_4(s_{ij})$ and $P_5(s_{ij})$ are of order $s^4$. At this order, $P_4(s_{ij})$ still has only a unique structure, while $P_5(s_{ij})$ have two independent structures, thus there are three independent parameters which can be reduced to two using the soft theorems of going from five points to four points. The six-point amplitude can be generally expressed in terms of a local polynomial term and terms containing factorization poles which are determined by lower-point and lower-dimensional amplitudes~\footnote{At this order only four-point amplitudes of order $s^2$ and $s^5$ contributes since order-$s^3$ four-point amplitude vanishes in $\mathcal{N}=4$ SYM.}. We find that the polynomial term has $13$ independent structures, and soft theorems can fix $10$ of them in terms of those of five-point amplitudes. However, clearly six-point amplitudes should be further constrained by supersymmetry, such as the SUSY Ward identity presented in the Appendix~\ref{Appendix:Ward}. We will discuss this more in the following section of computing mixed amplitudes with both dialtons and pions. 

\subsubsection{Dilaton and Pion mixed sectors} 

When R-symmetry pions involved, first of all at four and five points, the amplitudes are completely determined by the pure-dilaton amplitudes via maximal supersymmetry as shown in (\ref{eq:45pts}). As for higher-point amplitudes, due to the fact that the soft-pion theorems are only leading order the constraints are slightly less powerful. As we discussed, for mixed amplitudes with $n_1$ dilatons and $n_2$ R-symmetry pions the requisite bound for valid recursion is $n_1+{1\over 2} n_2 > k$ at order $s^{k}$. Thus at order $s^3$, all the amplitudes are fully determined (and again are two-loop exact), except the six-point amplitudes with pions only. Now, $\mathcal{N}=4$ SUSY imposes further constraints that help to completely fix these amplitudes. Let us study this exceptional case in details in what follows. 

As shown in details in Appendix \ref{Appendix:Ward}, we find that six-point SU(4)-violating component amplitudes must take the form, 
\bea
A(\phi_{12},\phi_{12},\phi_{34},\phi_{34},\phi_{34},\phi_{34}) = s_{12}^2\, P_6(s_{ij}) \,, 
\eea
where $P_6(s_{ij})$ is symmetric polynomials with six external legs. At order $s^3$, $P_6(s_{ij})$ is of order $s^1$ and vanishes due to the momentum conservation. Thus all the six-point amplitudes (with or without dilatons) can be expressed as linear combinations of SU(4)-preserving amplitudes, such as,
\bea
A(\phi_{12},\phi_{12},\phi_{12},\phi_{34},\phi_{34},\phi_{34}) \,.
\eea
A way of determining these amplitudes is to make an ansatz, and fix unknown parameters using soft theorems. For this particular case, the ansatz can be expressed as a factorization term with (two) four-point amplitudes $A_4(\phi_{12}, \phi_{12}, \phi_{34}, \phi_{34})$ as the residue (thus this term is two-loop exact), as well as a degree $s^3$ polynomial with $S_3 \times S_3$ symmetry which has $7$ independent structures. We find in fact in this case the soft-dilaton theorems alone are enough to determine the amplitudes, and the soft-pion theorems can serve as a consistent check. Explicitly, we find the amplitude to be given by
\bea
A(\phi_{12},\phi_{12},\phi_{12},\phi_{34},\phi_{34},\phi_{34})
&=& \left( gN^2 \over 32 \pi^2 m^4 \right)^2 \left[ {s_{12}^2 \, s_{56}^2 \over s_{124}} - 
 {1 \over 6} \left( s_{12}^3 + s_{45}^3 \right)   - 
 \left( s_{12}^2 s_{13} + s_{45}^2 s_{46} \right) \right. \cr
 &-& \left. {1 \over 3} \left( s_{12}s_{13}s_{23} + s_{45}s_{56}s_{46} \right) \right] 
 + \mathcal{P}_{\{123;456 \}} \, .
\eea
From this amplitude and similar ones with different R-symmetry indices, we can obtain all mixed amplitudes using the map in (\ref{eq:Rpions}). At order $s^4$, we find the same conclusion that with the help of the SUSY Ward identity one can fix all six-point amplitudes at this order, in terms of the four-point one, namely they are fully determined in terms of a single unknown coefficient $c^{(2)}_{4}(g,N)$. We then can apply soft-BCFW to determine all higher-point amplitudes, except a seven-point amplitude with six pions as well as an eight-point amplitude with eight pions. This obstruction can be understood by a simple large-$z$ counting. As we discussed in the previous section, supersymmetry should of course impose further constraints, and we believe they should eventually completely fix all amplitudes at this order in terms of the lowest-point one, especially given the fact that the pure-dilaton sector is fully determined. Similarly at order $s^5$, as far as for the constraints we have used, unlike the pure-dilaton amplitudes not all the mixed amplitudes can be determined in terms of the four-point one. As we discussed previously for pure-dilaton amplitudes at higher points, it is certainly of interest to explore systematically the SUSY Ward identity constraints, which has been very successfully applied to the ``MHV" higher-dimensional operators $F_-^2F_+^{2\ell}$ as well as SU(4)-breaking ones: $\phi^n F_-^2F_+^{2\ell}$. We will leave this investigation as a future research direction.

\subsection{6D $\mathcal{N}=(2,0)$ supersymmetry}
In $D=6$, the $\mathcal{N}=(2,0)$ theory contains a self-dual two-form and 5 scalars as its bosonic field content. It describes the theory of multiple M5-branes, and since it lacks a perturbative expansion parameter, it is a non-lagrangian theory. Moving on to the Coulomb branch provides such an expansion parameter. 
 
On the Coulomb branch 4 of the 5 scalars are R-symmetry Goldstone bosons of SO(5) $\rightarrow$ SO(4) and the remaining one is the dilaton. The generators of SO(4) $\sim$ SU(2) $\times$ SU(2) are conveniently represented using a pair of Grassmann odd variables  $(\eta_a,\tilde{\eta}_{a})$, with $a=1,2$ being a chiral spinor index of the SU(2) subgroup of the little group\footnote{Not to be confused with the SO(4) residual R-symmetry.} SO(4) $\subset$ SO(5,1):
\eq
\{J^+,J^z,J^-\}=\{\eta{\cdot}\eta,\;\;\eta{\cdot}\partial_{\eta}-1,\;\;\partial_\eta{\cdot}\partial_{\eta}\},\quad \{\tilde{J}^i\} =  \{{J}^i (\eta\rightarrow \tilde{\eta})\}
\eqe
where the inner products are defined via the contraction of the chiral spinor index, {\it i.e.} $\eta{\cdot}\eta\equiv\eta^a\eta_a = 2 \eta^1\eta^2$. Note that the additive constant $-1$ for $J^z$ is required by the commutator $[J^+,J^-]=J^z$. When the operator $J^z$ acts on the on-shell matrix elements one finds  
\eq
J^z\mathcal{A}_n\equiv\sum_{i}\left(\eta_i{\cdot}\partial_{\eta_i}-1\right)\mathcal{A}_n=0
\eqe
and similarly for $\tilde{J}^z$. As a result, the $n$-point amplitude turns out to be a polynomial of degree $(n,n)$ in the Grassmann variables. Following almost {\it verbatim} our discussion of $\mathcal{N}=4$ SYM in $D=4$, let us try to construct SUSY invariant local building blocks at four and five-points that are annihilated by the 16 susy operators $Q^{A+}=\lambda^{A}{\cdot}\eta$, $Q^{A-}=\lambda^{A}{\cdot}\partial_{\eta}$ and $\eta\rightarrow \tilde{\eta}$, with $A=1,\ldots 4$ a spinor index\footnote{In $D=6$ light-like momenta can be written as $P^{[AB]} = P^\mu \Gamma_\mu^{[AB]}  = \epsilon^{ab}\lambda_a^A \lambda_b^B$.} of SO$(5,1)$. The susy invariant four- and five- point amplitudes read 
\bea 
\mathcal{A}_4 &=& \delta^4(Q^+)\delta^4(\tilde Q^+) \sum_k P^{(k)}_4(s_{ij}) \, , \cr
\mathcal{A}_5 &=& \delta^4(Q^+)\delta^4(\tilde Q^+) \left({\sum_i \eta_i{\cdot}\tilde\eta_i} \right) \sum_k P^{(k)}_5(s_{ij})\, .
\eea
Acting with the derivative susy operators gives zero, since it generates terms that are proportional to the sum of total momentum or super-momentum, which vanish on the support of the delta functions. Thus following a similar analysis as in the D=4 case, the dilaton effective action is again completely fixed up to ten derivatives in terms of the three coefficients of the four-point operator.

\section{Scale vs Conformal symmetry} \label{section:scalevsconformal}

The relation between scale invariance and conformal invariance can also be studied for effective field theories (see e.g.~\cite{Nakayama:2013is} for a recent review). In our language, the question can be framed as follows: ``To what extent does the sub-leading soft theorem, due to broken conformal boost symmetry, follow from the leading behaviour stemming from broken dilation symmetry?'' First of all, we find that any five-point amplitude (which is a polynomial in $s_{ij}$) constrained by the leading soft theorem automatically satisfy the sub-leading soft theorem. This fact has been checked up to the very high $s^{11}$ order. For instance, at this particular order, four-point amplitudes involve two different polynomial structures, while five-point amplitudes depend on eleven parameters associated to as many independent polynomial structures. The leading soft theorem fixes two out of the eleven parameters in terms of the rest and those in the four-point amplitudes, and we find that the sub-leading soft theorem does not impose any further constraints.

More generally at higher points, according to the soft BCFW recursion relation, at order $s^{n}$, knowing the $2n$-point amplitude is enough to completely fix all the amplitudes with the same dimension by using the leading soft theorem alone, one may ask whether these amplitudes satisfy the sub-leading soft theorem automatically. Recall that from soft-BCFW recursion relations we have
\bea
A_{2n+1} = {1 \over 2 \pi i}  \oint_{{\cal C}_0} {dz \over z} {A_{2n+1}(z) \over F^{(1)}_{2n+1}(z)} \, . 
\eea
From the form of the $(2n{+}1)$-point in the soft BCFW representation, it is highly non-trivial that the amplitude also satisfies the sub-leading soft theorem. At order $s^2$, all the amplitudes are simply
\bea
A^{(2)}_n = c^{(2)}_n  ( s_{12}^2 + \mathcal{P}_n )  \, ,
\eea
and $c^{(2)}_n$ for $n>4$ are determined in terms of $c^{(2)}_{4}$ via the leading soft theorem. With such $c^{(2)}_n$ satisfying the leading soft theorem, in this relatively simple case one can show that $A^{(2)}_n$ also satisfies the sub-leading soft theorem. Beyond order $s^2$, the story becomes more interesting and non-trivial. We have checked explicitly for many non-trivial examples that this is indeed the case for amplitudes at orders $s^3, s^4$ and $s^5$. Let us take $s^3$ as an example to illustrate the idea. The inputs are the five-point amplitude at order $s^3$, 
\bea
A^{(3)}_5 =c^{(3)}_5 ( s_{12}^3 + \mathcal{P}_5 ) \, ,
\eea
as well as the four-point amplitude at order $s^2$, $A^{(2)}_4$. With these inputs one can construct the six-point amplitude using both leading and sub-leading soft theorems, and find, for instance in 4D, 
\bea 
A^{(3)}_6 &=& -c^{(3)}_5 ( s_{12}^3 + \mathcal{P}_6 ) - \left( {c^{(3)}_5 \over 2} + (c^{(2)}_4)^2 \right)( s_{123}^3 + \mathcal{P}_{6} ) \cr
&+& 
(c^{(2)}_4)^2 \left( (s_{12}^2 + s_{13}^2 + s_{23}^2){1 \over s_{123}} (s_{45}^2 + s_{46}^2 + s_{56}^2) + \mathcal{P}_{6} \right) \, . 
\eea
Now, the leading soft theorem alone allows us to determine $A^{(3)}_7$ in terms of lower-point and lower-derivative amplitudes. Explicitly, we find 
\bea \label{eq:A7s3}
A^{(3)}_7 =c^{(3)}_5 ( s_{12}^3 + \mathcal{P}_7 ) + \left( {c^{(3)}_5 } +3 {(c^{(2)}_4)^2 } \right)( s_{123}^3 + \mathcal{P}_7 ) 
-{ (c^{(2)}_4)^2 } A^{\rm fac}_7 \, ,
\eea
where $A_{\rm fac}$ is the factorization contribution, defined as
\bea
A^{\rm fac}_7 &=& (s_{12}^2 + s_{13}^2 + s_{23}^2) {1 \over s_{123}}
\left[ s_{45}^2 + s_{46}^2 + s_{47}^2 + 
   s_{56}^2 + s_{57}^2  + s_{67}^2+  (s_{47} + s_{57} +s_{67})^2  \right.  \cr
 &+& \left.
   (s_{45} + s_{46} + s_{47})^2 + (s_{45} + s_{56} + s_{57})^2 + (s_{46} + s_{56} + s_{67})^2
   \right] + \mathcal{P}_7 \, . 
\eea
It is then straightforward to verify that $A^{(3)}_7$ with particular parameters fixed by the leading soft theorem as in (\ref{eq:A7s3}) does satisfy the sub-leading soft theorem automatically. Similar construction or the use of recursion relations can be carried out for amplitudes of higher order, as we mentioned we have explicitly checked the statement up to order-$s^5$ local polynomial terms (namely due to the complication at this order, we set the factorization terms to vanish), which requires constructing the amplitudes until $10$ points using both leading and sub-leading soft theorems, and finally obtain the $11$-point amplitude using the leading soft theorem alone, and we find this amplitude does further satisfy the sub-leading soft theorem. 

\section{Conclusions} \label{section:conclusion}
In this paper, we initiate the systematic study of constraints on effective actions due to soft theorems of spontaneously broken symmetries where multiple GB modes are mixed under the broken symmetry. Using the one-loop and one-instanton effective action for $\mathcal{N}=4$ SYM in the Coulomb branch, we demonstrated the validity of the dilaton soft theorems as well as that of the newly derived R-symmetry pion soft theorems, both perturbatively and non-perturbatively. We have shown that with maximal susy, the dilaton effective action is completely determined up to ten derivatives in terms of two unknown coefficients parameterising the four-point amplitude. 

For CFTs which are non-Lagrangian, the dilaton effective actions are unique in the sense that the coefficients of the irrelevant operators are not functions of continuous parameter such as the coupling constant. However even with maximal SUSY, we've seen that broken and unbroken symmetries leave behind a large number of unknown coefficients. It is interesting to explore what are the other possible constraint that leads us to the unique action. An obvious possibility would be to explore the full implication of UV unitarity. At four points, this manifests itself as positivity constraint \cite{Adams:2006sv, Bellazzini:2015cra}.\footnote{Recently, using unitarity, analyticity and crossing symmetry, \cite{Bellazzini:2016xrt} shows that amplitudes that are softer than $s^2$ does not admit a non-trivial UV completion.} Needless to say that results beyond four points, while complicated, are desirable as this would be an alternative approach to gathering information on consistent CFTs.

In $D=4$, the maximal supersymmetric theory also enjoys S-duality at finite $N$. Furthermore in the large $N$ limit the UV theory on the Coulomb branch enjoys dual conformal symmetry~\cite{Drummond:2008vq}. It is thus a pressing question to understand to what extent does this input allow us to further fix the effective action. Also we have already discussed, it is important to have a better understanding of utilizing supersymmetry constraints at higher multiplicity, which would certainly reduce the independent parameters of higher-point amplitudes. In $D=3$, the massless degrees of freedom for the maximal theory are all Goldstone bosons. The eight scalars are identified as 7 Goldstone bosons from the breaking of SO(8) $R$-symmetry to SO(7), while the remaining one is the dilaton. Thus it would be interesting to explore the extent of uniqueness for its effective action when all broken and unbroken symmetry are taken into account. One may apply similar analysis to the low-energy expansion of string theories, since the string scattering amplitudes satisfy similar soft theorems, in particular the soft ``dilaton" theorems (for the closed-string dilaton)~\cite{DiVecchia:2015jaq}. 

We observed  and tested many highly non-trivial examples showing that amplitudes determined by recursion relations only based on the leading soft theorem satisfy the sub-leading soft theorem automatically. This observation leads to the supporting evidence that relativistic quantum field theories (under certain assumptions) with scale symmetry necessarily possess the enhanced conformal symmetry. It would certainly be interesting to study more on the possible equivalence between scale invariance and conformal invariance in the context of soft theorems.

Recently it was shown that the soft limit of Born-Infeld theory~\cite{Cachazo:2016njl}, at order $q^1$ in soft momentum is proportional to a larger theory involving the higher dimensional operators that mixes between the field strengths of Born-Infeld photons and  Yang-Mills gluons. Given that so far a majority of universal soft behaviours can be explained via symmetry, it will be interesting to study if there exists a hidden symmetry in the larger theory that would dictate such universal soft limits.

\vspace{0.4cm}

\section{Acknowledgements}

We are grateful to Nima Arkani-Hamed, Clay Cordova, Paolo Di Vecchia, Francesco Fucito, Raffaele Marotta, Francisco Morales, David Poland, Yassen Stanev and Xi Yin for discussions. We would also like to thank Oliver Schlotterer and Carlos Mafra for helping to carry out the analysis on 10D SYM integrands. 
Y-t.~H. and C-J. L.~are supported by MOST under the grant No.~103-2112-M-002-025-MY3. 

\appendix
\section{Dilaton vertices of one-instanton effective action} \label{Appendix:oneinstanton}

After preforming the fermionic $\theta$-integration and expanding in $1/v$, the one-instanton effective action in (\ref{seff0}) generates higher-dimensional vertices, from which we can read off scattering amplitudes of interest. Here we list some vertices involving dilatons that produce scattering amplitudes in (\ref{inst_amps}) of the section \ref{section:instantoncheck},
\begin{align}
v^8\Gamma^{(4)}[\varphi]&{=}(\de_\mu\de_\nu\varphi\,\de^\mu\de^\nu\varphi)^2{-}\,\de_\mu\de_\nu\varphi\,\de^\nu\de^\rho\varphi\,\de_\rho\de_\sigma\varphi\,\de^\sigma\de^\mu\varphi
\equiv (\de\de\varphi{\cdot}\de\de\varphi)^2{-}(\de\de\varphi{\cdot}\de\de\varphi{\cdot}\de\de\varphi{\cdot}\de\de\varphi)\\
v^9\Gamma^{(5)}[\varphi]&=-8\,\varphi(\de\de\varphi{\cdot}\de\de\varphi)^2+8\,\varphi(\de\de\varphi{\cdot}\de\de\varphi{\cdot}\de\de\varphi{\cdot}\de\de\varphi)\nonumber\\
&-8(\de\de\varphi{\cdot}\de\de\varphi)\,\de\varphi{\cdot}\de\de\varphi{\cdot}\de\varphi+8\,\de\varphi{\cdot}\de\de\varphi{\cdot}\de\de\varphi{\cdot}\de\de\varphi{\cdot}\de\varphi-2(\de\de\varphi{\cdot}\de\de\varphi{\cdot}\de\de\varphi \,\,\de\varphi{\cdot}\de\varphi)\\
v^{10}\Gamma^{(6)}[\varphi]&=36\,\varphi^2(\de\de\varphi{\cdot}\de\de\varphi)^2-36\,\varphi^2(\de\de\varphi{\cdot}\de\de\varphi{\cdot}\de\de\varphi{\cdot}\de\de\varphi)\nonumber\\
&+72\,\varphi(\de\de\varphi{\cdot}\de\de\varphi{\cdot}\de\de\varphi{\cdot}\de\de\varphi)\,\de\varphi{\cdot}\de\de\varphi{\cdot}\de\varphi-72\,\varphi\,\de\varphi{\cdot}\de\de\varphi{\cdot}\de\de\varphi{\cdot}\de\de\varphi{\cdot}\de\varphi\nonumber\\
&+18\,\varphi(\de\de\varphi{\cdot}\de\de\varphi{\cdot}\de\de\varphi)
+8(\de{\cdot}\de\de\varphi{\cdot}\de\varphi)^2-4\,\de\varphi{\cdot}\de\de\varphi{\cdot}\de\de\varphi{\cdot}\de\varphi+\frac{9}{2}(\de\de\varphi{\cdot}\de\de\varphi)\,(\de\varphi\de\varphi)^2\\
v^{11}\Gamma^{(7)}[\varphi]&=-120\,\varphi^3(\de\de\varphi{\cdot}\de\de\varphi)^2+120\,\varphi^3(\de\de\varphi{\cdot}\de\de\varphi{\cdot}\de\de\varphi{\cdot}\de\de\varphi)\nonumber\\
&-360\,\varphi^2(\de\de\varphi{\cdot}\de\de\varphi{\cdot}\de\de\varphi{\cdot}\de\de\varphi)\,\de\varphi{\cdot}\de\de\varphi{\cdot}\de\varphi+360\,\varphi^2\,\de\varphi{\cdot}\de\de\varphi{\cdot}\de\de\varphi{\cdot}\de\de\varphi{\cdot}\de\varphi\nonumber\\
&-90\,\varphi^2(\de\de\varphi{\cdot}\de\de\varphi{\cdot}\de\de\varphi)
-80\,\varphi(\de\varphi{\cdot}\de\de\varphi{\cdot}\de\varphi)^2+40\,\varphi\,\de\varphi{\cdot}\de\de\varphi{\cdot}\de\de\varphi{\cdot}\de\varphi\nonumber\\
&-45\,\varphi\,(\de\de\varphi{\cdot}\de\de\varphi)\,(\de\varphi\de\varphi)^2
-10\,\de\varphi{\cdot}\de\de\varphi{\cdot}\de\varphi\,(\de\varphi\de\varphi)^2\\
v^{12}\Gamma^{(8)}[\varphi]&=330\,\varphi^4(\de\de\varphi{\cdot}\de\de\varphi)^2-330\,\varphi^4(\de\de\varphi{\cdot}\de\de\varphi{\cdot}\de\de\varphi{\cdot}\de\de\varphi)\nonumber\\
&+1320\,\varphi^3(\de\de\varphi{\cdot}\de\de\varphi{\cdot}\de\de\varphi{\cdot}\de\de\varphi)\,\de\varphi{\cdot}\de\de\varphi{\cdot}\de\varphi-1320\,\varphi^3\,\de\varphi{\cdot}\de\de\varphi{\cdot}\de\de\varphi{\cdot}\de\de\varphi{\cdot}\de\varphi\nonumber\\
&+330\,\varphi^3(\de\de\varphi{\cdot}\de\de\varphi{\cdot}\de\de\varphi)
+440\,\varphi^2(\de{\cdot}\de\de\varphi{\cdot}\de\varphi)^2-220\,\varphi^2\,\de\varphi{\cdot}\de\de\varphi{\cdot}\de\de\varphi{\cdot}\de\varphi\nonumber\\
&+\frac{495}{2}\,\varphi^2\,(\de\de\varphi{\cdot}\de\de\varphi)\,(\de\varphi\de\varphi)^2
+110\,\varphi\,\de\varphi{\cdot}\de\de\varphi{\cdot}\de\varphi\,(\de\varphi\de\varphi)^2+\frac{15}{4}(\de\varphi\de\varphi)^4\\
v^{13}\Gamma^{(9)}[\varphi]&=-792\,\varphi^5(\de\de\varphi{\cdot}\de\de\varphi)^2+792\,\varphi^5(\de\de\varphi{\cdot}\de\de\varphi{\cdot}\de\de\varphi{\cdot}\de\de\varphi)\nonumber\\
&-3960\,\varphi^4(\de\de\varphi{\cdot}\de\de\varphi{\cdot}\de\de\varphi{\cdot}\de\de\varphi)\,\de\varphi{\cdot}\de\de\varphi{\cdot}\de\varphi+3960\,\varphi^4\,\de\varphi{\cdot}\de\de\varphi{\cdot}\de\de\varphi{\cdot}\de\de\varphi{\cdot}\de\varphi\nonumber\\
&-990\,\varphi^4(\de\de\varphi{\cdot}\de\de\varphi{\cdot}\de\de\varphi)
-1760\,\varphi^3(\de{\cdot}\de\de\varphi{\cdot}\de\varphi)^2+880\,\varphi^3\,\de\varphi{\cdot}\de\de\varphi{\cdot}\de\de\varphi{\cdot}\de\varphi\nonumber\\
&-990\,\varphi^3\,(\de\de\varphi{\cdot}\de\de\varphi)\,(\de\varphi\de\varphi)^2
-660\,\varphi^2\,\de\varphi{\cdot}\de\de\varphi{\cdot}\de\varphi\,(\de\varphi\de\varphi)^2-45\,\varphi\,(\de\varphi\de\varphi)^4.
\end{align}

\section{Sp(4) SUSY Ward identity} \label{Appendix:Ward}
Choosing the non-vanishing Sp(4) matrix elements to be $\Omega^{12}=\Omega^{34}=1$, one can have SU(4) violating amplitudes of the form 
\eq
A(\phi_{12},\phi_{12},\phi_{34},\phi_{34},\phi_{34},\phi_{34}) \, .
\eqe
They are represented in the following super amplitudes:
\eq
\mathcal{A}_6=\frac{\delta^8(Q)}{\langle 56\rangle^4}\frac{1}{[34]^4}\left(x_{1122}Y_{11}Y_{22}+x_{1212}Y_{12}Y_{12}\right)\,.
\eqe   
The coefficients $x_{1122}$ and $x_{1212}$ are linear combination of component amplitudes. Their explicit form will not be important here. $Y_{ij}=Y_{ji}$ and is given by 
\eqa
Y_{ij}&=&\left[([i 3]\eta^1_4+[ 34]\eta^1_i+[4i]\eta^1_3)([i 3]\eta^2_4+[ 34]\eta^2_i+[4i]\eta^2_3)\nonumber\right.\\
&+&\left[([i 3]\eta^3_4+[ 34]\eta^3_i+[4i]\eta^3_3)([i 3]\eta^4_4+[ 34]\eta^4_i+[4i]\eta^4_3) +(i\rightarrow j)\right]/2\,.
\eqae 
Note that it is a polynomial in pairs of $\eta_i^1\eta_j^2$ and $\eta_i^3\eta_j^4$. This is due to our choice of having  $\Omega^{12}=\Omega^{34}=1$. The component amplitude $A(\phi_{12},\phi_{12},\phi_{34},\phi_{34},\phi_{34},\phi_{34})$ comes from the coefficient of the polynomial $(\eta_5)^3(\eta_5)^4 (\eta_6)^3(\eta_6)^4(\eta_3)^3(\eta_4)^4(\eta_3)^3(\eta_4)^4(\eta_1)^1(\eta_1)^2(\eta_2)^1(\eta_2)^2$ in the super amplitude:
\eq
A(\phi_{12},\phi_{12},\phi_{34},\phi_{34},\phi_{34},\phi_{34})=\frac{\langle12\rangle^2[12]^2}{\langle56\rangle^2[34]^2}\left(x_{1122}-\frac{1}{2}x_{1212}\right)
\eqe
This can be compared to the $(\eta_5)^3(\eta_5)^4 (\eta_6)^3(\eta_6)^4(\eta_1)^3(\eta_1)^4(\eta_2)^3(\eta_2)^4(\eta_3)^1(\eta_3)^2(\eta_4)^1(\eta_4)^2$ coefficient:
\eq
A(\phi_{34},\phi_{34},\phi_{12},\phi_{12},\phi_{34},\phi_{34})=\frac{\langle34\rangle^2}{\langle56\rangle^2}\left(x_{1122}-\frac{1}{2}x_{1212}\right)
\eqe
Thus by supersymmetry arguments, we find that 
\eq
A(\phi_{12},\phi_{12},\phi_{34},\phi_{34},\phi_{34},\phi_{34})=\frac{s^2_{12}}{s^2_{34}}A(\phi_{34},\phi_{34},\phi_{12},\phi_{12},\phi_{34},\phi_{34})\,.
\eqe
The above identity shows that the amplitude $A(\phi_{12},\phi_{12},\phi_{34},\phi_{34},\phi_{34},\phi_{34})$ at any order must take the form, 
\bea
A(\phi_{12},\phi_{12},\phi_{34},\phi_{34},\phi_{34},\phi_{34}) = s_{12}^2\, \mathcal{P}_6(s_{ij}) \,, 
\eea
and $\mathcal{P}_6(s_{ij})$ has the full $S_6$ permutation symmetry. Furthermore, it is easy to see that $A(\phi_{12},\phi_{12},\phi_{34},\phi_{34},\phi_{34},\phi_{34})$ cannot have any factorization poles, thus $\mathcal{P}_6(s_{ij})$ can mostly have a pole of $1/s_{12}^2$, but due to the permutation symmetry such a pole is not allowed. So in conclusion, $\mathcal{P}_6(s_{ij})$ is a symmetric polynomial in $s_{ij}$, whose classification is much simpler now. Similar analysis applies to other six-point Sp(4) amplitudes, and the same conclusion can be reached.

\end{document}